# Switching management in couplers with biharmonic longitudinal modulation of refractive index

Yaroslav V. Kartashov[1] and Victor A. Vysloukh[2]

[1]*ICFO-Institut de Ciencies Fotoniques, and Universitat Politecnica de Catalunya, Mediterranean Technology Park, 08860 Castelldefels (Barcelona), Spain*

[2]*Departamento de Fisica y Matematicas, Universidad de las Americas – Puebla, Santa Catarina Martir, 72820, Puebla, Mexico*

We address light propagation in couplers with longitudinal biharmonic modulation of refractive index in neighboring channels. While simplest single-frequency out-of-phase modulation allows suppression of coupling for strictly defined set of resonant frequencies, the addition of modulation on multiple frequency dramatically modifies the structure of resonances. Thus, additional modulation on double frequency may suppress primary resonance, while modulation on triple frequency causes fusion of primary and secondary resonances.

*OCIS codes: 190.0190, 190.6135*

The optical materials with inhomogeneous refractive index landscapes provide exceptional opportunities for diffraction control, especially when the refractive index vary both in the transverse plane and in the direction of propagation of radiation [1,2]. In such structures discrete diffraction-managed solitons exist [3-5], while the propagation direction and transverse shape of the beam can be dynamically altered [6-9]. Among the most spectacular phenomena in longitudinally modulated structures [10-21] are resonant suppression of light coupling (analogous to coherent destruction of tunneling in double-well structures perturbed by a harmonic driving force [17]) and dynamic localization. Such phenomena are possible in waveguide arrays [10-16] and two-channel structures [17-21] when channels bend periodically in the propagation direction [10-12,18,19] or when the width/refractive index in neighboring channels oscillate out-of-phase [13-16,21]. Up to date only simplest harmonic laws of longitudinal modulation were considered. Only recently stacking segments with different curvatures in periodically bending structures were used to observe polychromatic dynamic localization [22,23]. The suppression of coupling for narrow excitations in systems with multiple-frequency oscillations of refractive index was not addressed.

In this Letter we consider light propagation in directional couplers with complex out-of-phase "biharmonic" modulation of refractive index in neighboring channels. We show that addition of even weak modulation on a multiple frequency dramatically modifies basic structure of resonances and allows management of positions and strengths of resonances.

The propagation of laser radiation along the $\xi$-axis of directional coupler is described by the Schrödinger equation for the dimensionless field amplitude $q$:

$$i\frac{\partial q}{\partial \xi} = -\frac{1}{2}\frac{\partial^2 q}{\partial \eta^2} - pR(\eta,\xi)q. \qquad (1)$$

Here $\eta$ and $\xi$ are the normalized transverse and longitudinal coordinates, while $p$ stands for the refractive index modulation depth. The refractive index distribution is given by $R(\eta,\xi) = [1 + Q(\xi)]\exp[-(\eta + w_s/2)^6/w_\eta^6] + [1 - Q(\xi)]\exp[-(\eta - w_s/2)^6/w_\eta^6]$, where $w_s$ is the separation between channels and $w_\eta$ is the channel width. The out-of-phase longitudinal refractive index modulation in the channels is described by $Q(\xi) = \mu\sin(\Omega\xi) + \alpha\sin(n\Omega\xi)$, where $\mu$ is the depth of modulation on main frequency $\Omega$, while $\alpha$ is the depth of modulation on frequency with multiplicity $n$. We set $w_\eta = 0.3$, $w_s = 3.2$, and $p = 2.78$. For characteristic transverse scale 10 $\mu$m and wavelength $\lambda = 800$ nm this describes two 3 $\mu$m waveguides with 32 $\mu$m separation and refractive index depth $\sim 3.1\times 10^{-4}$.

To study beam dynamics we excited left channel at $\xi = 0$ using linear guided mode of isolated waveguide. When $\mu = \alpha = 0$ the light periodically switches between channels with beating frequency $\Omega_b = 2\pi/T_b$, where $T_b = 100$ for our parameters. When $\mu \neq 0$, $\alpha = 0$ the coupling suppression can be achieved for strictly defined set of modulation frequencies [21]. This is accompanied by resonant increase of distance-averaged energy flow $U_m = L^{-1}\int_0^L d\xi \int_{-\infty}^0 |q(\eta,\xi)|^2 d\eta \Big/ \int_{-\infty}^0 |q(\eta,0)|^2 d\eta$ trapped in the input channel (here $L$ is the final distance). The primary resonance in the dependence $U_m(\Omega)$ has largest frequency $\Omega_{rp}$, while frequencies of secondary resonances $\approx \Omega_{rp}/n$, where $n = 2,3,4...$. For moderate $\mu$ values $\Omega_{rp} \sim \mu$, while for $\mu \to 0$ one has $\Omega_{rp} \to \Omega_b/2$.

The central result of this Letter is that addition of even weak modulation on multiple frequency drastically modifies the basic resonance structure. Thus, Fig. 1(a) shows dependencies $U_m(\Omega)$ around primary resonance for increasing $\alpha$ values and $n = 2$. The additional modulation on double frequency gradually destroys primary resonance and resonant value $U_{mr}$ of averaged energy that was $\approx 1$ at $\alpha = 0$ drops down to $1/2$ already for $\alpha$ values much smaller than $\mu$ [Fig. 1(b)]. The frequency $\Omega_{rp}$ remains almost unchanged with $\alpha$.

The localization decreases not only in primary resonance, but also in secondary resonances, although larger amplitudes $\alpha$ are necessary for their destruction. The critical value $\alpha_{\rm cr}$ at which $U_{\rm mr}$ decreases down to $1/2$ monotonically grows with increase of $\mu$, but in all cases $\alpha_{\rm cr} \ll \mu$.

The picture is completely different when the main-frequency modulation is combined with that on triple frequency. This results in visible increase of strengths of resonances and also leads to progressive shift in the frequencies of primary and secondary resonances that at $n=3$ occurs in such way that primary resonance frequency $\Omega_{\rm rp}$ decreases, while secondary resonance frequency $\Omega_{\rm rs}$ increases resulting in gradual fusion of these resonances [Fig. 1(c)]. The width of such combined resonance is much larger than widths of isolated resonances at $\alpha=0$, i.e. almost perfect coupling suppression occurs for broader band of modulation frequencies. The broadening of resonances may be potentially employed to achieve coupling suppression also for polychromatic radiation. Dependencies of $\Omega_{\rm rp}$, $\Omega_{\rm rs}$, and the difference $\delta\Omega = \Omega_{\rm rp} - \Omega_{\rm rs}$ on amplitude $\alpha$ are shown in Fig. 1(d). The critical amplitude $\alpha_{\rm cr}$ at which $\delta\Omega$ decreases down to $\Omega_{\max}/2$, where $\Omega_{\max} = \delta\Omega_{\alpha=0}$, is a monotonically increasing function of $\mu$. Higher-order resonances also may fuse although in that case resonant frequencies may depend on $\alpha$ in a complex way.

The phase mismatch $\phi$ between modulations on different frequencies that may be introduced into function $Q(\xi) = \mu\sin(\Omega\xi) + \alpha\sin(n\Omega\xi + \phi)$ notably modifies coupling dynamics. Thus, for $n=2$, $\mu=0.20$, $\alpha=0.03$ the averaged energy in principal resonance grows from $U_{\rm m}(\phi=0)=0.69$ to $U_{\rm m}(\phi=\pi/2)=0.98$, and then decreases to $U_{\rm m}(\phi=\pi)=0.74$. A similar trend was found for $n=3$. This indicates importance of concrete functional shape of longitudinal modulation $Q(\xi)$ which strongly depends on phase shift.

Nontrivial features of reshaping of resonance curves may be interpreted in the frames of tight-binding approximation operating with field amplitudes in adjacent channels $q_1(\xi)$, $q_2(\xi)$ whose dynamics is described by the coupled ordinary differential equations [21]. Introducing symmetric and antisymmetric modes $C_{\rm s} = (q_1+q_2)/2$, $C_{\rm a} = (q_1-q_2)/2$, one gets:

$$\begin{aligned}\frac{dC_{\rm s}}{d\xi} &= +i\kappa\,{\rm Re}[F(\xi)]C_{\rm s} - \kappa\,{\rm Im}[F(\xi)]C_{\rm a},\\ \frac{dC_{\rm a}}{d\xi} &= -i\kappa\,{\rm Re}[F(\xi)]C_{\rm a} + \kappa\,{\rm Im}[F(\xi)]C_{\rm s}.\end{aligned} \quad (2)$$

Here $\kappa = p\int_{-\infty}^{\infty} w_1 w_2 R_{\xi=0} d\eta$ is the coupling constant, $w_{1,2}$ describe profiles of guided modes in two waveguides and are normalized in such way that $\int_{-\infty}^{\infty} w_{1,2}^2 d\eta = 1$, the old notations $\mu$ and $\alpha$ were kept for amplitudes of longitudinal modulation in tight-binding approximation (they are in fact given by $\mu p \int_{-\infty}^{\infty} w_1^2 R_{\xi=0} d\eta$ and $\alpha p \int_{-\infty}^{\infty} w_1^2 R_{\xi=0} d\eta$), while the real and imaginary parts of the factor $F = \exp[(2i\mu/\Omega)\cos(\Omega\xi) + (2i\alpha/n\Omega)\cos(n\Omega\xi)]$ determine phase shifts of symmetric and antisymmetric modes and their coupling. In the absence of refractive index modulation $\text{Re}(F) = 1$ and $\text{Im}(F) = 0$. Single-channel excitation corresponds to $C_s(0) = C_a(0) = 1$ and switching occurs at the distance $L$ where accumulated phase difference $\arg[C_s(L)] - \arg[C_a(L)] = \pi$. Importantly, in the presence of refractive index modulation mode dephasing and mode coupling is suppressed if $\text{Re}(F) \to 0$ and $\text{Im}(F) \to 0$. Figure 2 shows real (curves 1) and imaginary (curves 2) parts of distance-averaged factor $F_{\text{av}} = L^{-1} \int_0^L F(\xi) d\xi$ versus $\Omega$ in comparison with $U_m(\Omega)$ dependence (curves 3). At $\alpha = 0$ the value $\text{Im}(F_{\text{av}})$ is negligible and positions of resonances are defined by zeros of $\text{Re}(F_{\text{av}})$ [Fig. 2(a)]. The modulation on double frequency results in increase of $|\text{Im}(F_{\text{av}})|$ within the band of primary resonance and gradually destroys this resonance [Fig. 2(b)]. Additional modulation on triple frequency shifts $\text{Re}(F_{\text{av}})$ curve upwards that diminishes the interval between zeros of $\text{Re}(F_{\text{av}})$ and causes fusion of primary and secondary resonances [Fig. 2(c)].

Typical propagation scenarios in couplers with biharmonic modulation are presented in Figs. 3 and 4. While in the presence of longitudinal modulation with only frequency the periodic energy exchange [Fig. 3(a)] between channels of coupler may be almost completely suppressed in resonance [Fig. 3(b)], the addition of modulation with $n = 2$ destroys localization as long as $\alpha$ grows [Figs. 3(c) and 3(d)]. In contrast, addition of modulation with $n = 3$ leading to shift and broadening of resonances may result in suppression of coupling [Figs. 4(b) and 4(d)] at modulation frequencies that were well outside any resonances in system with $\mu \neq 0$, $\alpha = 0$ [Figs. 4(a) and 4(c)]. The combination of main modulation with modulation at much higher frequencies with $n = 4, 5, ...$ is not effective because such high-frequency oscillations average out and one has to increase amplitude $\alpha$ considerably in order to notably affect resonance structure - something that leads also to growth of radiative losses.

Summarizing, we showed that complex biharmonic longitudinal modulation of refractive index in channels of directional coupler allows management of positions and strengths of resonances where coupling suppression is achieved. Such management can be also realized in more complicated guiding structures such as waveguide arrays.

# References with titles

# References without titles

# Figure captions

Figure 1. (a) $U_{\rm m}$ versus $\Omega$ at $\alpha=0$ (curve 1), $\alpha=0.025$ (curve 2), and $\alpha=0.040$ (curve 3) for $\mu=0.2$, $n=2$. (b) $U_{\rm m}$ versus $\alpha$ at $\Omega=\Omega_{\rm rp}$ for $\mu=0.15$ (curve 1) and $\mu=0.30$ (curve 2) at $n=2$. (c) $U_{\rm m}$ versus $\Omega$ for $\alpha=0$ (curve 1) and $\alpha=0.11$ (curve 2) for $\mu=0.15$, $n=3$. (d) $\Omega_{\rm rp}$, $\Omega_{\rm rs}$, and $\delta\Omega=\Omega_{\rm rp}-\Omega_{\rm rs}$ versus $\alpha$ for $\mu=0.1$, $n=3$.

Figure 2. $\text{Re}(F_{\rm av})$ (curves 1), $\text{Im}(F_{\rm av})$ (curves 2) and $U_{\rm m}$ (curves 3) versus $\Omega$ for (a) $\alpha=0.00$, $n=0$, (b) $\alpha=0.40$, $n=2$, and (c) $\alpha=0.15$, $n=3$, at $\mu=0.258$.

Figure 3. Propagation dynamics in (a) unmodulated coupler at $\mu=0$, $\alpha=0$, and in modulated couplers at (b) $\mu=0.2$, $\alpha=0$, (c) $\mu=0.2$, $\alpha=0.025$, and (d) $\mu=0.2$, $\alpha=0.05$. Here $n=2$, $L=4T_{\rm b}$. Modulation frequency $\Omega=3.45\Omega_{\rm b}$ in (b)-(d) corresponds to $\Omega_{\rm rp}$ for $\mu=0.2$, $\alpha=0$.

Figure 4. Propagation dynamics in modulated couplers with (a) $\alpha=0.00$, $\Omega=1.78\Omega_{\rm b}$, (b) $\alpha=0.11$, $\Omega=1.78\Omega_{\rm b}$, (c) $\alpha=0.00$, $\Omega=2.11\Omega_{\rm b}$, and (d) $\alpha=0.11$, $\Omega=2.11\Omega_{\rm b}$. Here $\mu=0.15$, $n=3$, $L=4T_{\rm b}$.

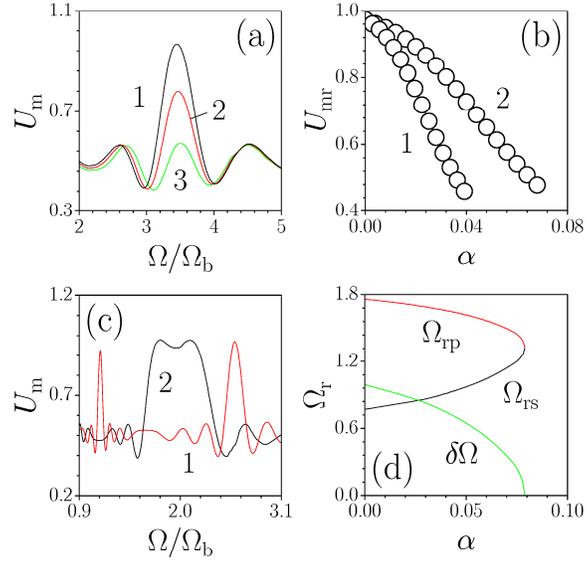

Figure 1. (a) $U_m$ versus $\Omega$ at $\alpha = 0$ (curve 1), $\alpha = 0.025$ (curve 2), and $\alpha = 0.040$ (curve 3) for $\mu = 0.2$, $n = 2$. (b) $U_m$ versus $\alpha$ at $\Omega = \Omega_{rp}$ for $\mu = 0.15$ (curve 1) and $\mu = 0.30$ (curve 2) at $n = 2$. (c) $U_m$ versus $\Omega$ for $\alpha = 0$ (curve 1) and $\alpha = 0.11$ (curve 2) for $\mu = 0.15$, $n = 3$. (d) $\Omega_{rp}$, $\Omega_{rs}$, and $\delta\Omega = \Omega_{rp} - \Omega_{rs}$ versus $\alpha$ for $\mu = 0.1$, $n = 3$.

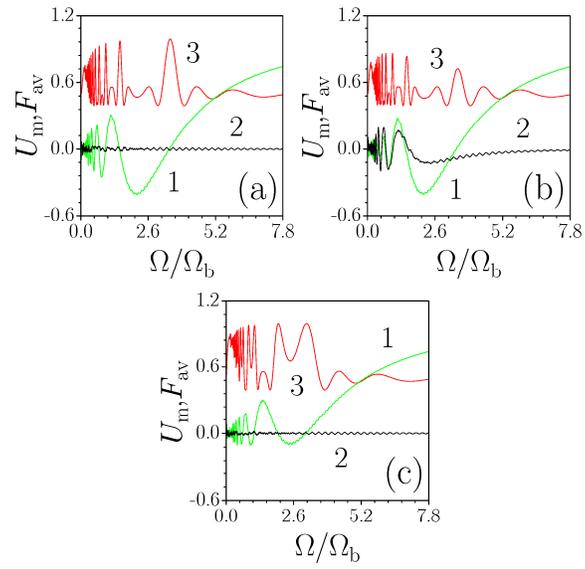

Figure 2. Re($F_{\mathrm{av}}$) (curves 1), Im($F_{\mathrm{av}}$) (curves 2) and $U_{\mathrm{m}}$ (curves 3) versus $\Omega$ for (a) $\alpha = 0.00$, $n = 0$, (b) $\alpha = 0.40$, $n = 2$, and (c) $\alpha = 0.15$, $n = 3$, at $\mu = 0.258$.

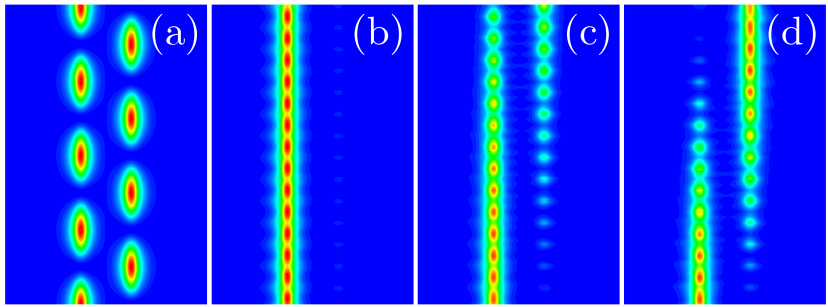

Figure 3. Propagation dynamics in (a) unmodulated coupler at $\mu = 0$, $\alpha = 0$, and in modulated couplers at (b) $\mu = 0.2$, $\alpha = 0$, (c) $\mu = 0.2$, $\alpha = 0.025$, and (d) $\mu = 0.2$, $\alpha = 0.05$. Here $n = 2$, $L = 4T_{\rm b}$. Modulation frequency $\Omega = 3.45\Omega_{\rm b}$ in (b)-(d) corresponds to $\Omega_{\rm rp}$ for $\mu = 0.2$, $\alpha = 0$.

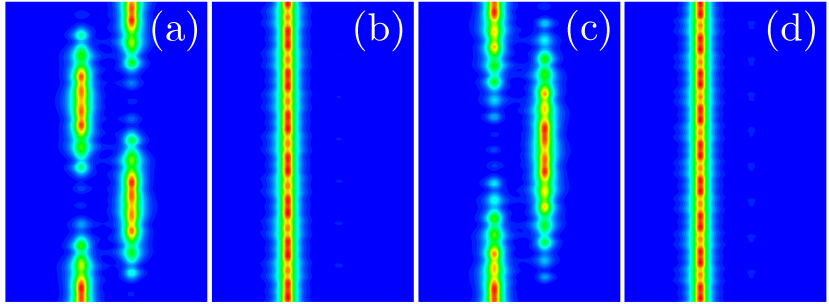

Figure 4. Propagation dynamics in modulated couplers with (a) $\alpha = 0.00$, $\Omega = 1.78\Omega_{\rm b}$, (b) $\alpha = 0.11$, $\Omega = 1.78\Omega_{\rm b}$, (c) $\alpha = 0.00$, $\Omega = 2.11\Omega_{\rm b}$, and (d) $\alpha = 0.11$, $\Omega = 2.11\Omega_{\rm b}$. Here $\mu = 0.15$, $n = 3$, $L = 4T_{\rm b}$.